\documentclass[aps,superscriptaddress,twocolumn,showpacs,showkeys,amsmath,amssymb]{revtex4}

\usepackage{graphicx}
\usepackage{dcolumn}
\usepackage{epstopdf} 

\begin{document}


\title{Cost of $s$-fold Decisions in Exact Maxwell-Boltzmann, \\ Bose-Einstein and Fermi-Dirac Statistics}

\author{Robert K. Niven}
\email{r.niven@adfa.edu.au}
\affiliation{
School of Aerospace, Civil and Mechanical Engineering, \\ The University of New South Wales at ADFA, \\Northcott Drive, Canberra, ACT, 2600, Australia
}



\date{September 30, 2005}

\begin{abstract}
The exact forms of the degenerate Maxwell-Boltzmann (MB), Bose-Einstein (BE) and Fermi-Dirac (FD) entropy functions, derived by Boltzmann's principle without the Stirling approximation (Niven, Physics Letters A, 342(4) (2005) 286), are further examined.  Firstly, an apparent paradox in quantisation effects is resolved using the Laplace-Jaynes interpretation of probability.  The energy cost of learning that a system, distributed over $s$ equiprobable states, is in one such state (an ``$s$-fold decision") is then calculated for each statistic. The analysis confirms that the cost depends on one's knowledge of the number of entities $N$ and (for BE and FD statistics) the degeneracy, extending the findings of Niven (2005).
\end{abstract}

\pacs{
02.50.Cw, 
03.65.Ta, 
05.20.-y, 
05.30.-d 
}

\keywords{entropy, information theory, combinatorial, statistical mechanics, measurement problem, Maxwell's demon}

\maketitle

\section{Introduction}

The combinatorial definition of entropy of Boltzmann \cite{Boltzmann} and Planck \cite{Planck} stands as one of the most profound equations of human discovery. Recently, it was used to derive the exact forms of the degenerate Maxwell-Boltzmann (MB), Bose-Einstein (BE) and Fermi-Dirac (FD) entropy functions, without the Stirling approximation (i.e. the total number of entities, $N$, and/or the number of entities $n_i$ in each state $i$, and/or the degeneracy do not necessarily approach infinity) \cite{Niven}.  The exact entropy functions are supersets of their corresponding Stirling-approximate forms \cite{Boseetc,Tolman,Davidson,Kapur}, but are distinct from other supersets such as the Tsallis entropy \cite{Tsallis}.  The analysis shows that the concept of ÒinformationÓ not only encompasses knowledge of the realization (complexion) of a system, but also knowledge of $N$, and for BE and FD systems, knowledge of the degeneracy of each state.  

The energy cost (in bits) of a ``binary decision" - i.e. the cost of learning that a system, distributed over two equiprobable states, is in one such state - was calculated in each statistic.  It was found that a binary decision costs $<1$ bit in MB and BE systems, and in a narrow range within FD systems, if one has additional knowledge \cite{Niven}.  In BE systems, a zero cost is theoretically attainable.  However, the cost is $>1$ bit in BE and FD systems in the absence of such knowledge, except in the Stirling limits.  In consequence, the observation of an individual BE or FD entity is thermodynamically irreversible (requires the input of energy or information); this explains the need to destroy a boson or fermion in order to observe it, and hence  ``the collapse of the wavefunction" \cite{Niven}.
	
Exact MB, BE and FD statistics are further examined here.  Firstly, an apparent paradox relating to quantisation effects is resolved.  The previous analysis is then extended to consider an ``$s$-fold decision", i.e. the cost of learning that a system, distributed over $s$ equiprobable states, is in one such state.  The findings generalise those reported previously \cite{Niven}.  As before, the discussion draws upon the combinatorial definition of entropy \cite{Boltzmann,Planck,Boseetc,Tolman,Davidson,Kapur}; the equivalence of information, energy and negative entropy \cite{Szilard,Shannon,Brillouin,Leff}; and the information-theoretic understanding of the second law of thermodynamics \cite{Szilard,Brillouin,Leff}.

\section{Background}
%

The Boltzmann \cite{Boltzmann} - Planck \cite{Planck} combinatorial definition of entropy can be expressed as:
\begin{equation}
H = \frac{{\ln \mathbb{W}}}{N} + C
\label{eq1}
\end{equation}
where $H$ is a dimensionless entropy per unit entity; $\mathbb{W}$ is the statistical weight or number of possible realizations (complexions) of the system, of equal probability; $N$ is the number of entities (e.g., atoms or molecules); and $C$ is an arbitrary constant (reference datum).  The weights for degenerate MB, BE and FD statistics are \cite{Boseetc,Tolman,Davidson,Kapur}:
\begin{equation}
\mathbb{W}_{MB}  = N!\prod\limits_{i = 1}^s {\frac{{g_i ^{n_i } }}{{n_i !}}}  = N!\prod\limits_{i = 1}^s {\frac{{(\alpha _i N)^{p_i N} }}{{(p_i N)!}}}
\label{eq2}
\end{equation}
\begin{equation}
\mathbb{W}_{BE}  = \prod\limits_{i = 1}^s {\frac{{(g_i  + n_i  - 1)!}}{{(g_i  - 1)!n_i !}}}  = \prod\limits_{i = 1}^s {\frac{{(\alpha _i N + p_i N - 1)!}}{{(\alpha _i N - 1)!(p_i N)!}}} 
\label{eq3}
\end{equation}
\begin{equation}
\mathbb{W}_{FD}  = \prod\limits_{i = 1}^s {\frac{{g_i !}}{{n_i !(g_i  - n_i )!}}}  = \prod\limits_{i = 1}^s {\frac{{(\alpha _i N)!}}{{(p_i N)!(\alpha _i N - p_i N)!}}}
\label{eq4}
\end{equation}
where $i$ denotes each distinguishable state, from a total of $s$ distinguishable states; $n_i$ is the number of entities in each state $i$; $g_i$ is the degeneracy (multiplicity) of each state $i$; $p_i = n_i/N$ is the probability of an entity being in state $i$; and $\alpha_i=g_i/N$ is the relative degeneracy of state $i$.  As noted \cite{Niven}, it is mathematically convenient to consider the MB weight as a function of $g_i$, and the BE and FD weights as functions of $\alpha_i$.

From Eqs. (\ref{eq1}-\ref{eq4}), the exact entropy functions are \cite{Niven}:
\begin{equation}
H_{MB}^x  = \frac{1}{N} \sum\limits_{i = 1}^s {\left\{ { -\ln [(p_i N)!] + p_i \ln [N!] + p_i N \ln g_i } \right\}}
\label{eq5}
\end{equation}
\begin{equation}
H_{BE}^x  = \frac{1}{N} \sum\limits_{i = 1}^s \ln \frac{(\alpha _i N + p_i N - 1)!}{(\alpha _i N - 1)!(p_i N)!} 
\label{eq6}
\end{equation}
\begin{equation}
H_{FD}^x  = \frac{1}{N} \sum\limits_{i = 1}^s \ln \frac{(\alpha _i N)!}{(p_i N)!(\alpha _i N - p_i N)!}
\label{eq7}
\end{equation}
where superscript $x$ implies the exact entropy.  Here we take $C=0$ in each case, and also bring the external $N!$ of the MB weight (Eq. \ref{eq2}) inside the sum using the natural constraint $\sum\nolimits_{i = 1}^s {p_i  = 1} $.

Applying the Stirling approximation to each factorial ($\ln x! \to x\ln x - x$ as $x \to \infty$), Eqs. (\ref{eq5}-\ref{eq7}) reduce to their Stirling-approximate forms \cite{Boseetc,Tolman,Davidson,Kapur}:
\begin{equation}
H_{MB}^{St}  \approx  - \sum\limits_{i = 1}^s {p_i \ln \frac{{p_i }}{{g_i }}} 
\label{eq8}
\end{equation}
\begin{equation}
H_{BE}^{St}  \approx \sum\limits_{i = 1}^s {\left[ {(\alpha _i  + p_i )\ln (\alpha _i  + p_i ) - \alpha _i \ln \alpha _i  - p_i \ln p_i } \right]} 
\label{eq9}
\end{equation}
\begin{equation}
H_{FD}^{St}  \approx \sum\limits_{i = 1}^s {\left[ { - (\alpha _i  - p_i )\ln (\alpha _i  - p_i ) + \alpha _i \ln \alpha _i  - p_i \ln p_i } \right]} 
\label{eq10}
\end{equation}

\section{An Apparent Paradox}

Before proceeding, it is worth drawing out an apparent paradox in the implementation of ``exact statistical mechanics" - not spelt out previously \cite{Niven} - which in fact is not a paradox.  Consider a non-degenerate (i.e. $g_i=1$) MB system, for example a die with $s$ faces rolled $N$ times, from which the data are collected without regard to order.  For $s=2$ this reduces to the simple case of a coin tossed $N$ times.  In the Stirling limits $N \to \infty$ and $n_i  \to \infty,\forall i$, if the die is ÒtrueÓ (difficult to manufacture in practice), we have no difficulty with the statement that $p_i=s^{-1}, \forall i$; or for our coin, $p_1 = p_2 = \frac{1}{2}$.  Alternatively, if a trickster can roll a die or toss a coin so that it always lands with face $j$ up, we assign $p_i=1$ for $i=j$ and $p_i=0$ otherwise.

What about the other extreme, at $N=1$?  Consider two situations:

\noindent (a)	If we have already rolled the die, and completely know the outcome, we can assign $n_i=1,i=j; n_i=0,i \ne j$ for some $j$, hence $p_i=1,i=j; p_i= 0,i \ne j$. Thus in a situation of complete knowledge, the numbers of entities in each state - and hence the probabilities - are {\it quantised}.  For $N>1$, this quantisation will generate discrete fractional values of $p_i$.

\noindent (b)	On the other hand, if we do not know the outcome (e.g. before we toss the die), but consider that the die is true, we have no choice except to make the probability assignments $p_i=s^{-1},\forall i$, using the principle of insufficient reason \cite{Jaynes}, irrespective of the fact that the outcome will be quantised. If we convert these into numbers of entities within each state, we obtain $n_i=s^{-1},\forall i$. Similarly, for $N>1$ we obtain $n_i=N/s,\forall i$.  In other words, when we have no knowledge of the realization, the probability and entity assignments {\it override} the quantisation inherent in the problem.  This does not mean that we have a simultaneous superposition of all faces uppermost on our die, nor a superposition of heads and tails on our coin (nor indeed, a superpositional alive-dead cat \cite{Schrod}); it only means that our description is based on an imperfect state of knowledge.  There is no need to demand that this description be physically realizable.

Thus at low $N$, the analysis relies upon the Laplace-Jaynes interpretation of probabilities as ``plausibilities", or assignments based on what is known, rather than as measurable frequencies \cite{Jaynes,Jaynes2}.  The author recognises that this view causes great difficulty for many physicists, but it is the only rational method with which to assign probabilities.  Readers who do not understand this approach are urged to thoroughly read Jaynes.  Of course, in the Stirling limits $N\to\infty$ and $n_i\to\infty,\forall i$, both the ``before" and ``after" scenarios can be physically realized, and can thus be conceptualised as measurable frequencies.

\section{Effect of an \lowercase{$s$}-fold Decision}

Now that the apparent paradox is understood, consider an ``$s$-fold decision" as defined earlier, which encompasses the above die problem.  Before the decision, for equiprobable states we assign $p_i=s^{-1},\forall i$. After the decision, we find that the system occupies only one state, whence $p_i=1,i=j; p_i=0,i \ne j$.  Certainly, with increasing $N$ and $s$ we will be more and more surprised by this outcome, hence the information gained will also increase.  The energy cost of the decision is \cite{Szilard,Shannon,Brillouin,Niven}:
\begin{eqnarray}
\Delta E =-\Delta I = H_{final}-H_{init}~~(``nats")\nonumber \\
=S_{final}-S_{init}~~(``{\it s}{\rm{-}}bits")   
\label{eq11}
\end{eqnarray}
where $\Delta I$ is the information gain, $H_{init}$ and $H_{final}$ are the initial and final entropies (from Eqs. (\ref{eq5}-\ref{eq10})), and $S=H/\ln s$ is a rescaled entropy.  The units of ``nats" reflect entropies based on the natural logarithm; however, it is more convenient to use ``logical" logarithms of base 2 for binary decisions (``bits"), base 3 for ternary decisions, etc; whence base $s$ for $s$-fold decisions (``$s$-bits").  These are adopted in the last form of Eq. \ref{eq11}, using
$\log _s x = \ln x/\ln s$.

For finite $N$, provided we know $N$, we can use the exact form of $S_{init}$ in Eq. (\ref{eq11}).  However, if we don't know $N$, we must (by default) adopt the Stirling-approximate form of $S_{init}$ as the reference datum, since it reflects a state of higher entropy (lower information).  This leads to two separate costs:
(a) the cost of learning $N$, given by $\Delta E^N =-(S_{init}^x-S_{init}^{St})$; and
(b) the cost of learning the realization, given by $\Delta E^x =-(S_{final}^x-S_{init}^x)$.  
The sum $\Delta E^T=\Delta E^N+\Delta E^x =-(S_{final}^x-S_{init}^{St})$
gives the total cost, in $s$-bits, of learning both $N$ and the realization. 

By definition, an $s$-fold decision provides 1 $s$-bit of information about some observable system.  The cost $\Delta E$ must however be ``paid for" from some external reservoir of negative entropy or information \cite{Szilard,Brillouin,Leff}.  Thus in calculating $\Delta E^x$ or $\Delta E^T$, we are particularly interested in processes for which $\Delta E<1$ or $>1$ $s$-bit, since these respectively produce a net information surplus or deficit in the combined observable-reservoir system, in apparent defiance of the second law of thermodynamics \cite{Brillouin,Leff}.

\section{Calculations}

The initial and final entropies and energy cost for an $s$-fold decision in each statistic, for both Stirling-approximate and exact cases, are listed in Table \ref{tab:table1} ($\Delta E^N$ and $\Delta E^T$ are not listed but readily calculated).  For convenience, it is assumed that $g_i=g, \alpha_i = \alpha,\forall i$, and $x! = \Gamma (x+1)$ for non-integer $x$.  The resulting entropy surfaces - or the energy costs - can be plotted against their governing variables, to reveal their effect.  In the following, although the plots are drawn as continuous in $N$ and $s$, it is understood that both are discrete; furthermore, all plots commence at the physically meaningful $N=1$ and $s=2$.
	
\begin{table*}
\caption{\label{tab:table1}Initial and final entropies and energy cost of an $s$-fold decision, for Stirling-approximate and exact statistics}
\begin{ruledtabular}
\begin{tabular}{ccccc}
Name&$S_{init}$&$S_{final}$&$\Delta E^{St}$ or $\Delta E^x$\\ 
 &($s$-bits)&($s$-bits)&($s$-bits)\\
\hline
\multicolumn{4}{l}{Stirling-approximate form}\\
MB&$\log _s g + 1$&$\log _s g$&1\\
BE&$\log _s \frac{{(\alpha s + 1)^{\alpha s + 1} }}{{(\alpha s)^{\alpha s} }}$&$\log _s \frac{{(\alpha  + 1)^{\alpha  + 1} }}{{\alpha ^\alpha  }}$&$\log _s \frac{{\alpha ^\alpha  (\alpha s + 1)^{\alpha s + 1} }}{{(\alpha  + 1)^{\alpha  + 1} (\alpha s)^{\alpha s} }} \le 1$\\
FD&$\log _s \frac{{(\alpha s)^{\alpha s} }}{{(\alpha s - 1)^{\alpha s - 1} }}$&$\log _s \frac{{\alpha ^\alpha  }}{{(\alpha  - 1)^{\alpha  - 1} }}$&$\log _s \frac{{(\alpha  - 1)^{\alpha  - 1} (\alpha s)^{\alpha s} }}{{\alpha ^\alpha  (\alpha s - 1)^{\alpha s - 1} }} \ge 1$\\
\hline
\multicolumn{4}{l}{Exact form}\\
MB&$\frac{1}{N}\log _s \frac{{N!}}{{\left( {\frac{N}{s}!} \right)^s }} + \log _s g$&$\log _s g$&$\frac{1}{N}\log _s \frac{{N!}}{{\left( {\frac{N}{s}!} \right)^s }}$\\
BE&$\frac{s}{N}\log _s \frac{{\left( {\alpha N + \frac{N}{s} - 1} \right)!}}{{(\alpha N - 1)!\frac{N}{s}!}}$&$\frac{1}{N}\log _s \frac{{(\alpha N + N - 1)!}}{{(\alpha N - 1)!N!}}$&$\frac{1}{N}\log _s \frac{{N!}}{{\left( {\frac{N}{s}!} \right)^s }}\frac{{\left[ {\left( {\alpha N + \frac{N}{s} - 1} \right)!} \right]^s }}{{(\alpha N + N - 1)!\left[ {(\alpha N - 1)!} \right]^{s - 1} }}$\\
FD&$\frac{s}{N}\log _s \frac{{(\alpha N)!}}{{\left( {\alpha N - \frac{N}{s}} \right)!\frac{N}{s}!}}$&$\frac{1}{N}\log _s \frac{{(\alpha N)!}}{{(\alpha N - N)!N!}}$&$\frac{1}{N}\log _s \frac{{N!}}{{\left( {\frac{N}{s}!} \right)^s }}\frac{{[(\alpha N)!]^{s - 1} (\alpha N - N)!}}{{\left[ {\left( {\alpha N - \frac{N}{s}} \right)!} \right]^s }}$\\
\end{tabular}
\end{ruledtabular}
\end{table*}

\begin{figure}
\includegraphics[width=80mm]{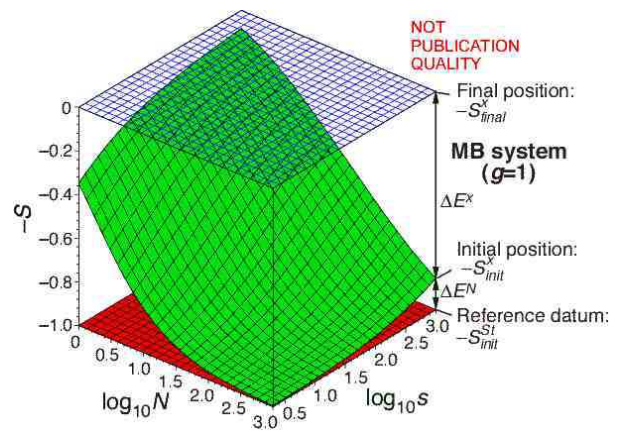}
\caption{Information-energy diagram for an exact MB system for $g=1$}
\label{fig:MBg1}
\end{figure}

For a non-degenerate MB system ($g=1$), the reference, initial and final entropies are plotted (in a negative sense) against $N$ and $s$ in Figure \ref{fig:MBg1}.   As shown, in this statistic the total cost is always 1 $s$-bit.  In the Stirling limit $N \to \infty$, $\Delta E^N \to 0$, but the value of $N$ above which the Stirling approximation can be considered valid increases with increasing $s$.  At finite $N$, if one has knowledge of $N$, the decision can be completed for $\Delta E^x<1$ $s$-bit.  Furthermore, this cost diminishes with increasing s (in the limit $s \to \infty$, $\Delta E^x \to 0$).
  
\begin{figure*}
\begin{center}
\setlength{\unitlength}{0.6pt}
  \begin{picture}(700,300)
   \put(0,0){\includegraphics[width=65mm]{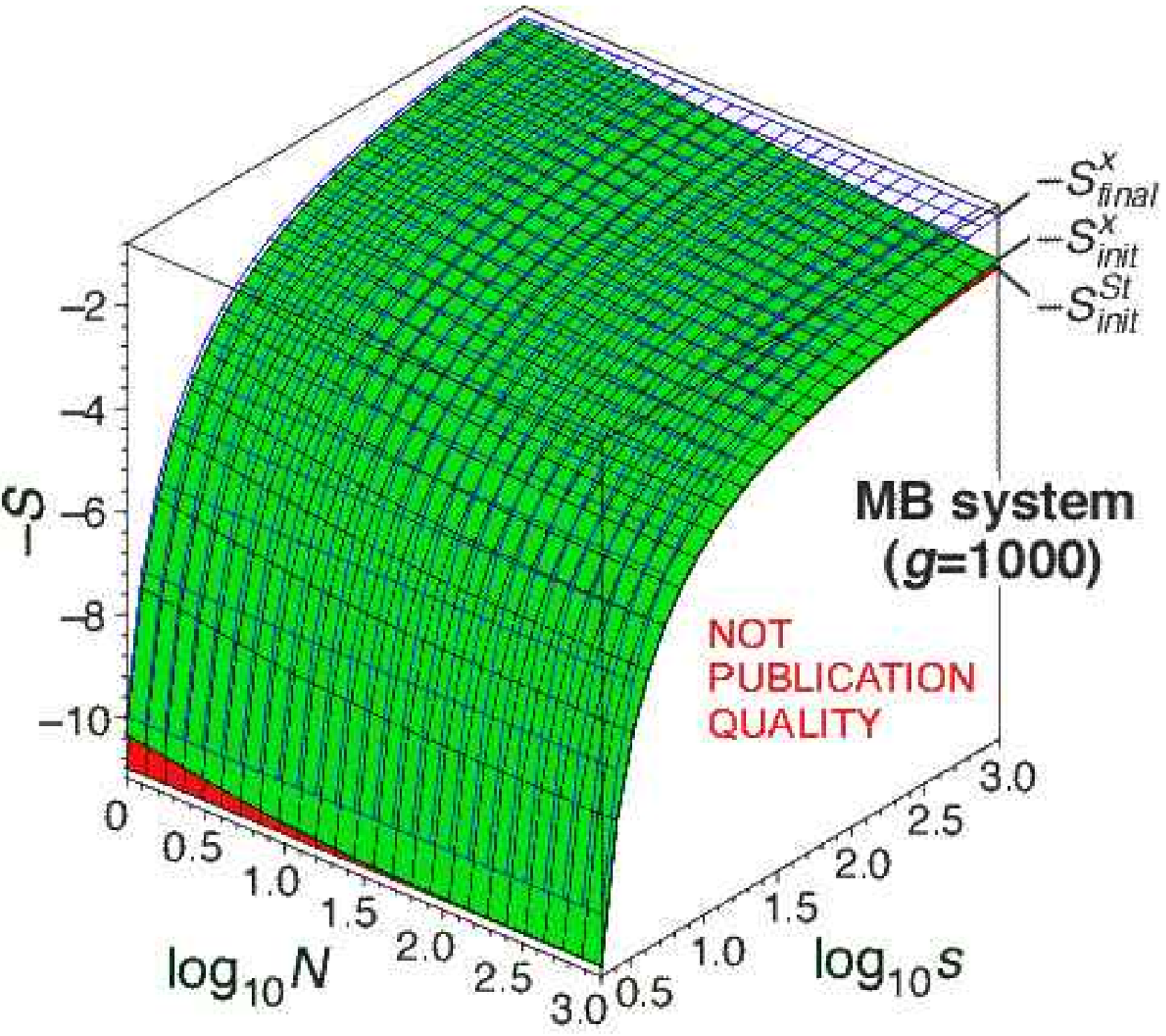} }
   \put(0,0){\small (a)}
   \put(350,0){\includegraphics[width=65mm]{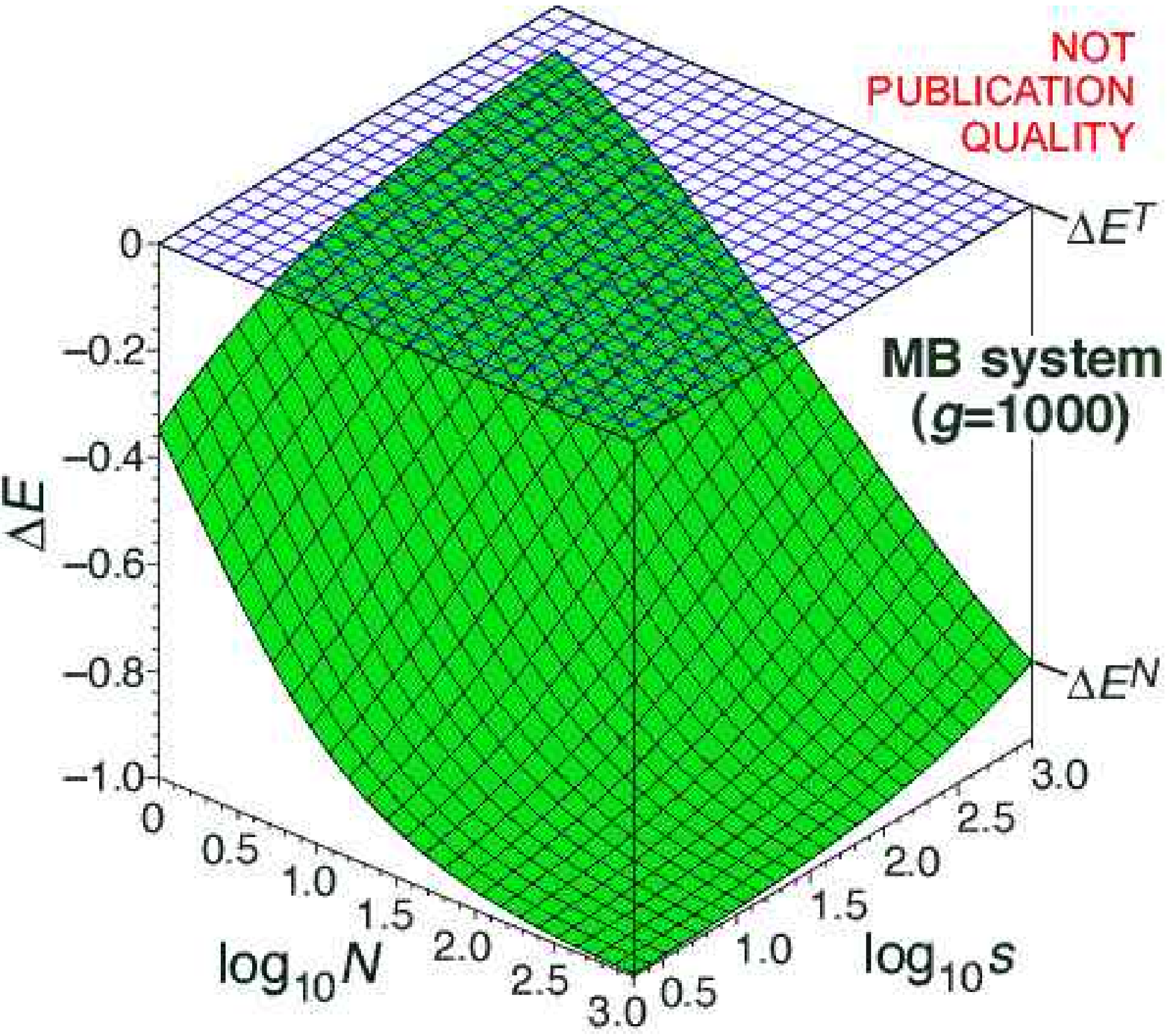} }
   \put(350,0){\small (b)}
  \end{picture}
\end{center}
\caption{Information-energy diagrams for exact MB systems, for $g=1000$, showing (a) entropy surfaces, and (b) net energy costs.}
\label{fig:MBg1000}
\end{figure*}

A plot of the various entropy functions for a degenerate MB system with $g=1000$ (Figure \ref{fig:MBg1000}a) reveals the effect of degeneracy.  However, if we plot the energy costs (Figure \ref{fig:MBg1000}b), we obtain surfaces which are identical to those in Figure \ref{fig:MBg1}.  In other words, for exact MB statistics, the magnitude of the degeneracy has no effect on the net cost of an $s$-fold decision, regardless of $N$ or $s$.

\begin{figure*}
\begin{center}
\setlength{\unitlength}{0.6pt}
  \begin{picture}(700,600)
   \put(0,0){\includegraphics[width=65mm]{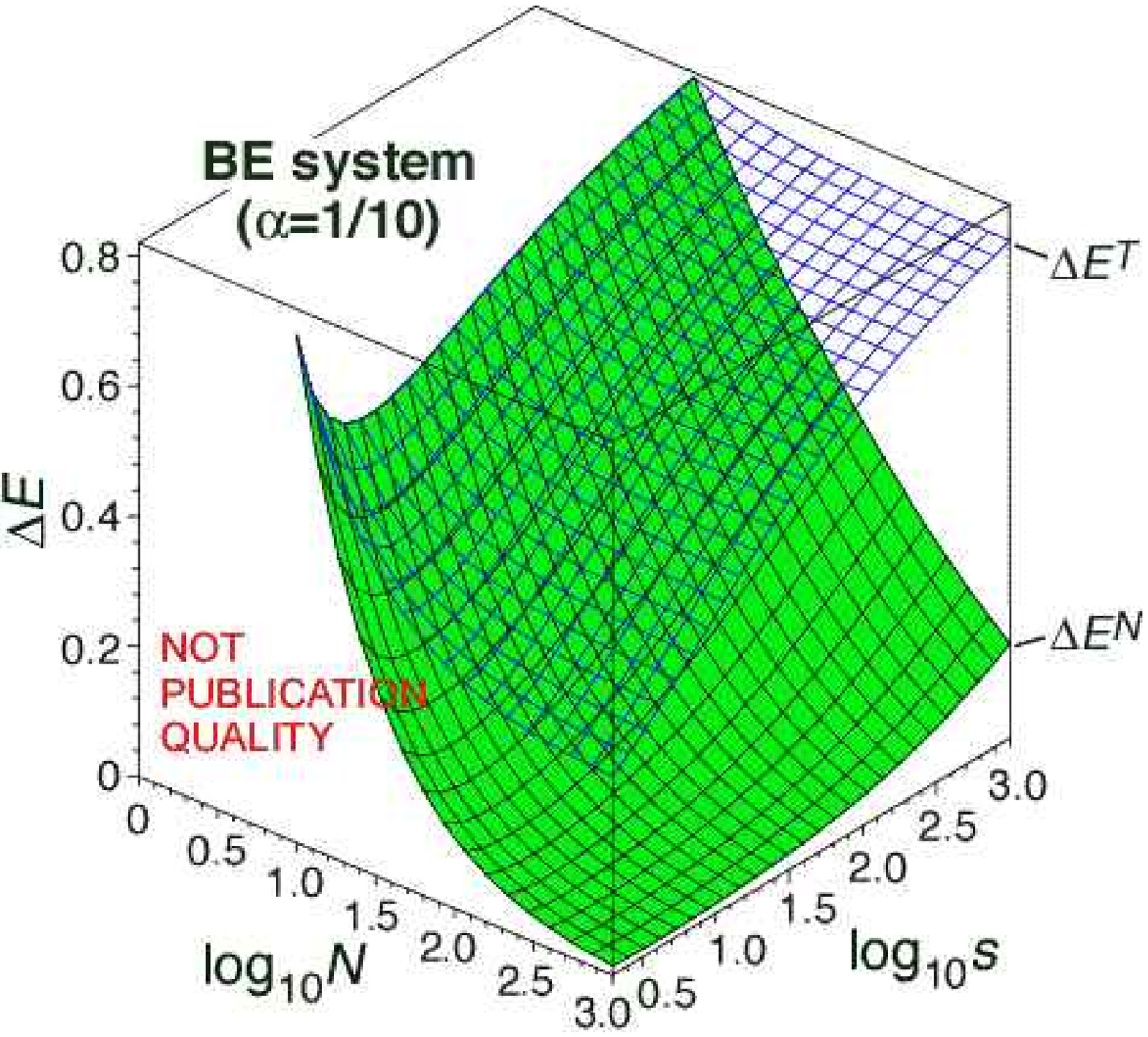} }
   \put(0,0){\small (c)}
   \put(350,0){\includegraphics[width=65mm]{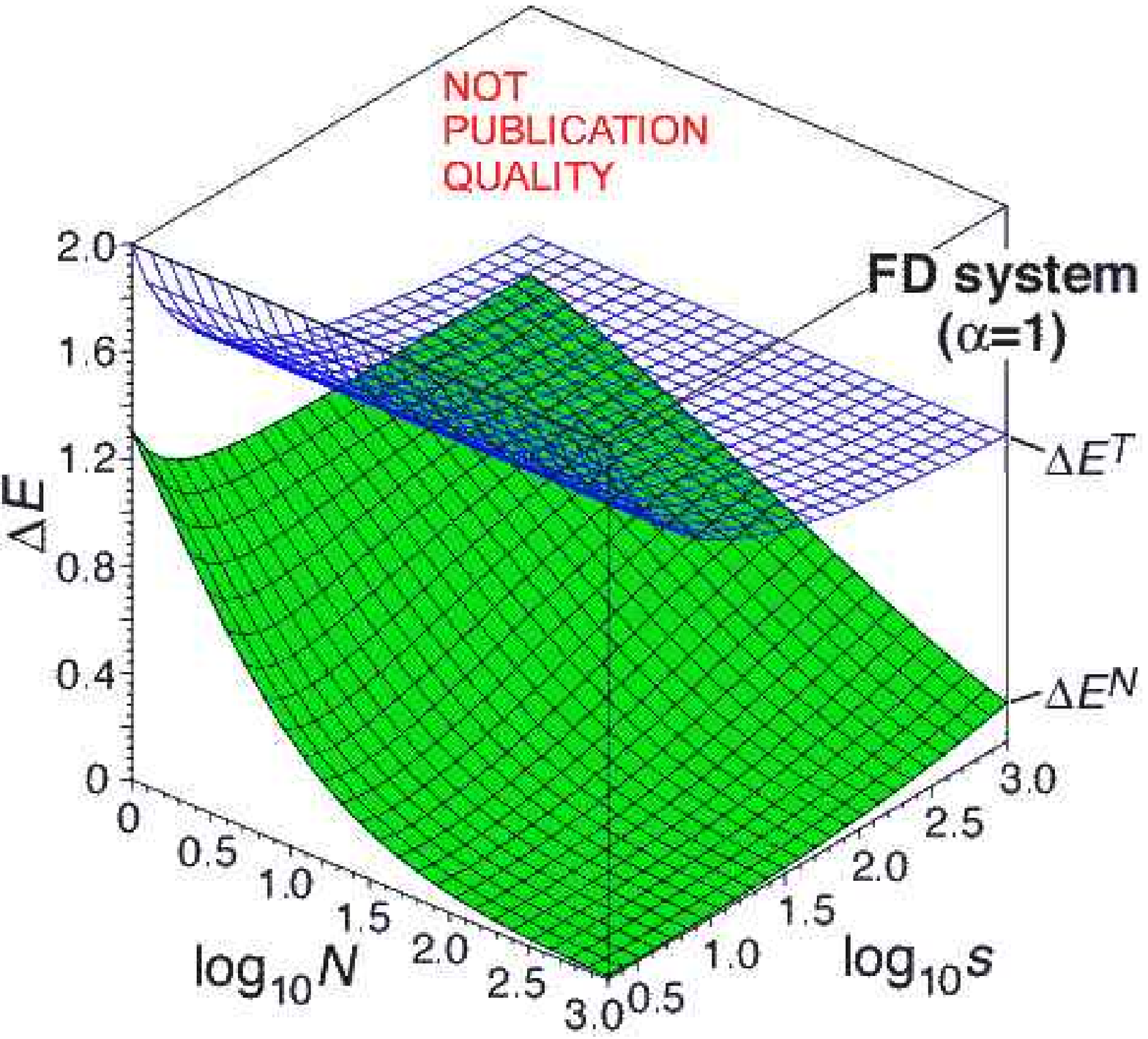} }
   \put(350,0){\small (d)}
   \put(0,300){\includegraphics[width=65mm]{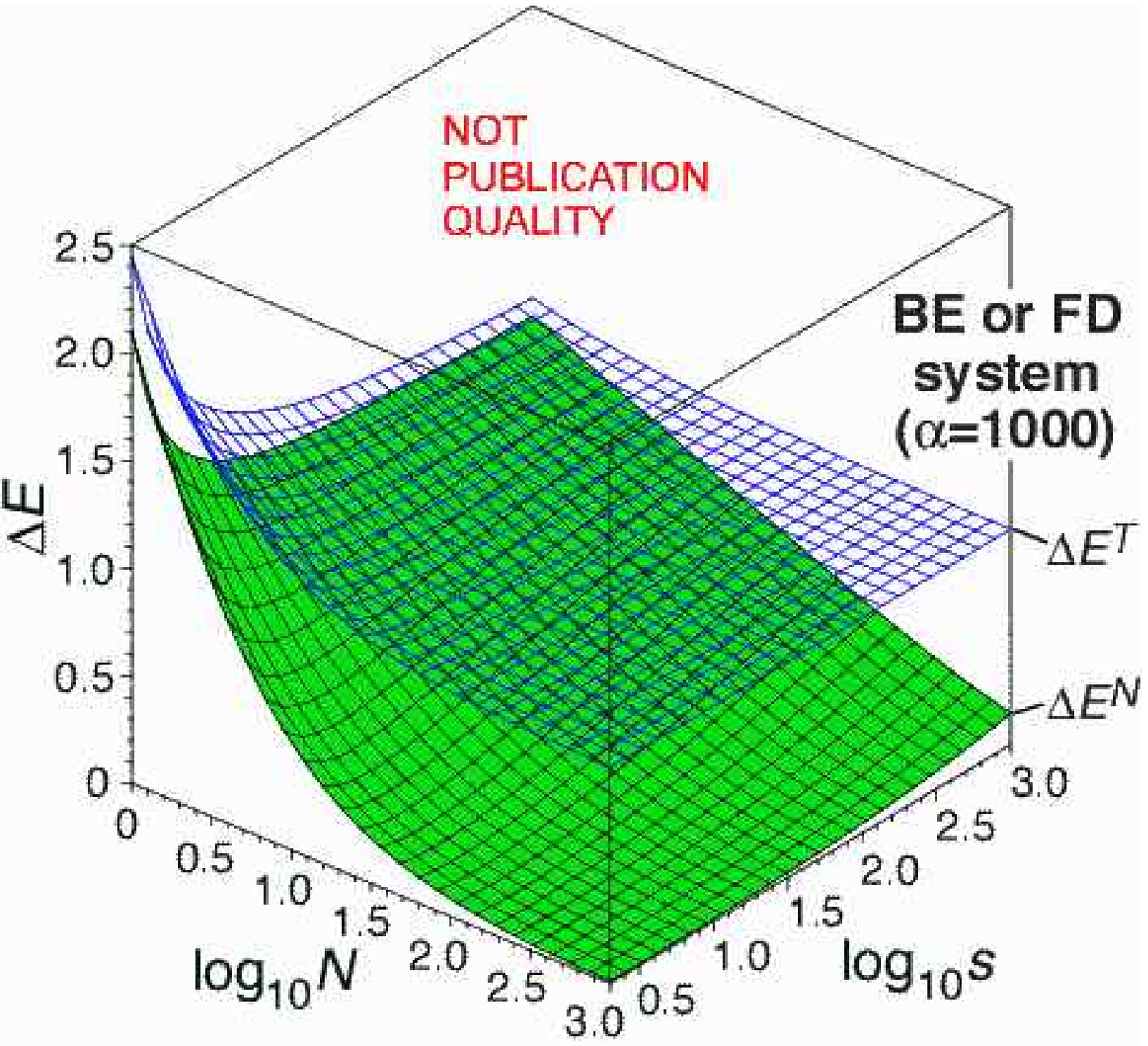}}
   \put(0,300){\small (a)}
   \put(350,300){\includegraphics[width=65mm]{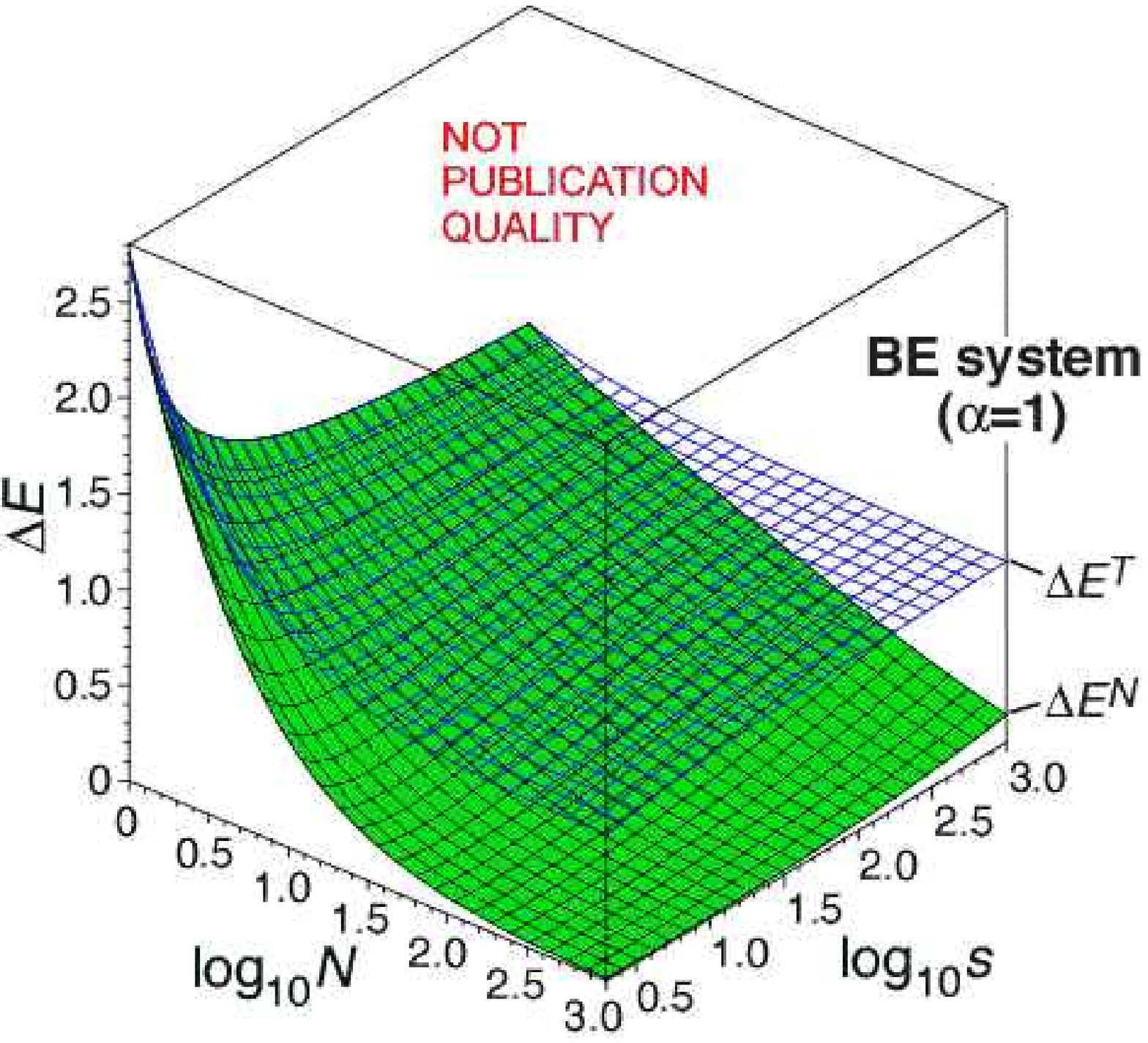}}
   \put(350,300){\small (b)}
  \end{picture}
\end{center}
\caption{Information-energy diagrams of energy costs in exact BE and FD systems, for (a) $\alpha=1000$ ($\approx \alpha \to \infty$); (b) exact BE system, $\alpha=1$; (c) exact BE system, $\alpha=\frac{1}{10}$; and (d) exact FD system, $\alpha=1$.}
\label{fig:BEFD}
\end{figure*}

The energy costs for an exact BE or FD system at $\alpha=1000$ (which very closely approximates $\alpha \to \infty$ \cite{Niven}) are plotted in Figure \ref{fig:BEFD}a (the actual entropies, not shown, exhibit ``folded" surfaces somewhat similar to Figure \ref{fig:MBg1000}a).  As evident, for $N \to \infty$ the cost of an $s$-fold decision is 1 $s$-bit, but once again the $N$ above which the Stirling approximation is valid increases with increasing $s$.  However, in contrast to the MB statistic, for finite $N$ the cost surfaces curve upwards, such that $\Delta E^N$ increases significantly as $N \to 1$, even as $\Delta E^x$ decreases.  The net effect is that with knowledge of $N \ne \infty$, an $s$-fold decision can be obtained for $\Delta E^x<1$ $s$-bit; however, without this knowledge, the total cost is $\Delta E^T>1$ $s$-bit, except in the Stirling limit ($N \to \infty$). 

In the limit as $\alpha \to \infty$, the various initial and final BE and FD entropies all become infinite; however, their differences approach the following values:
\begin{equation}
\Delta E_{BE}^x  = \Delta E_{FD}^x  \approx \frac{1}{N}\log _s \frac{{N!}}{{\left({\frac{N}{s}!} \right)^s}}
\label{eq12}
\end{equation}
\begin{equation}
\Delta E_{BE}^N  = \Delta E_{FD}^N  \approx \frac{1}{N}\log _s \left[ {\left( {\frac{N}{s}!} \right)^s e^N \left( {\frac{s}{N}} \right)^N } \right]
\label{eq13}
\end{equation}
\begin{equation}
\Delta E_{BE}^T  = \Delta E_{FD}^T  \approx \frac{1}{N}\log _s \left[ {e^N \left( {\frac{s}{N}} \right)^N N!} \right]
\label{eq14}
\end{equation}
Plots of the latter two functions are essentially identical to those in Figure \ref{fig:BEFD}a.

The effect of degeneracy on the BE statistic is illustrated in Figures \ref{fig:BEFD}b-c, respectively for $\alpha=1$ and $\alpha=\frac{1}{10}$.  In either case, if both $N$ and $\alpha \ne \infty$ are known, it possible to complete the $s$-fold decision for $\Delta E^x<1$ $s$-bit, even in the Stirling limit ($N \to \infty$).  Indeed, a zero cost is theoretically attainable in either system at $N=1/\alpha$, independent of $s$ (for $N<1/\alpha$, the cost surfaces cross to produce $\Delta E^x<0$; however, this occurs at a fractional degeneracy $g<1$, and is thus non-physical).  If only $\alpha \ne \infty$ is known, the total cost is higher, but there exists a range of $N$ and $s$ for which $\Delta E^T<1$.
  
For the FD statistic, the effect of degeneracy is shown in Figure \ref{fig:BEFD}d for $\alpha=1$.  As found previously \cite{Niven}, in FD statistics, knowledge of $\alpha$ {\it increases} the cost of an $s$-fold decision (for $N \to \infty$ and $s=2$, the cost is exactly 2 bits).  This effect does however diminish with increasing $s$.  FD systems are therefore much less favourable to the generation of an information surplus, although there exists a range of $N$ and $s$ for which $\Delta E^x<1$ or $\Delta E^T<1$.

\section{Conclusions}
The above analysis extends the previous finding \cite{Niven} that ``information" about a physical system is connected not only with knowledge of its realization, but also with knowledge of $N$, and in the case of BE and FD statistics, with knowledge of $\alpha$.  MB systems are again found to be ``well behaved", in that the total cost is always equal to 1 $s$-bit, regardless of $N$ and $s$, whereas BE and FD do not satisfy this condition.  With knowledge of $N$ and/or the degeneracy, it is possible to achieve an $s$-fold decision for $\Delta E^x<1$ $s$-bit in all systems examined (for FD systems, with some restrictions on $N$ and $s$).  Without knowledge of $N$ or degeneracy, the total cost is always $\Delta E^T>1$ $s$-bit in BE and FD systems, but approaches 1 $s$-bit in the Stirling limit $N \to \infty$.  The findings generalise those reported previously \cite{Niven}.  


\begin{acknowledgments}
The author thanks Cliff Woodward for questioning the apparent paradox, and Ra Inta and participants at the 2005 NEXT Sigma-Phi conference, Crete, Greece, for interesting discussions.
\end{acknowledgments}



\end{document}